\documentclass[twocolumn]{aa}  
\pdfoutput=1
\usepackage[utf8]{inputenc}
\PassOptionsToPackage{hyphens}{url}
\usepackage[english]{babel}
\usepackage{natbib}
\usepackage{graphicx}
\usepackage{amsmath}
\usepackage[varg]{txfonts}
\usepackage{setspace}
\usepackage{epstopdf}
\usepackage{tabularx}
\usepackage{ulem}
\usepackage{hyperref}

\newcommand{\fsequa}{\, \, \, .}
\newcommand{\comequa}{\, \, \, ,}

\newcommand{\Msol}{\mathit{M}_{\odot}}

\newcommand{\au}{\rm{au}}

\newcommand{\bef}{\begin{figure}[!t]}
\newcommand{\eef}{\end{figure}}

\begin{document} 

\institute{Hamburger Sternwarte, Universit\"at Hamburg, Gojenbergsweg  112, 21029 Hamburg, Germany \label{inst1}\\ \email{vperdelwitz@hs.uni-hamburg.de}}
\title{A new approach to distant solar system object detection in large survey data sets}

\titlerunning{New approach to SSO detection}

\author{V. Perdelwitz \and M. V\"olschow \and H.\,M. M\"uller}

\offprints{{\tt volker.perdelwitz@hs.uni-hamburg.de}}

 \date{Received: 7 November 2017 / Accepted: 19 April 2018}

\abstract
  {The recently postulated existence of a giant ninth planet in our solar system has sparked search efforts for distant solar system objects (SSOs) both via new observations and archival data analysis. Due to the likely faintness of the object in the optical and infrared regime, it has so far eluded detection.} 
  {We set out to re-analyze data acquired by the {\it Wide-Field Infrared Survey Explorer} (WISE), an all-sky survey well suited for the detection of SSOs.}
  {We present a new approach to SSO detection via parallactic fitting. Using the heliocentric distance as a fit parameter, our code transforms groups of three or more single observation point sources to heliocentric coordinates under the assumption that all data stem from an object. The fact that the orbit of a distant SSO is approximately linear in heliocentric coordinates over long time-scales can be utilized to produce candidates, which can then be confirmed with follow-up observations.}
  {We demonstrate the feasibility of the approach by {\it a posteriori} detecting the outer SSO Makemake within WISE data.
An all-sky search for Planet Nine yielded no detection.}
  {While the postulated Planet Nine eluded detection by our algorithm, we tentatively predict that this new approach to moving-object analysis will enable the discovery of new distant SSOs that cannot be discovered by other algorithms. Especially in cases of sparse data observed over long time spans, our approach is unique and robust due to the use of only one fit parameter.}


\maketitle 

\section{Introduction}
\label{sec:intro}
The last few decades have seen great advances in the understanding of the outer solar system. Seventy-four years after the discovery of Pluto \citep{1946ASPL....5...73T}, \cite{2004ApJ...617..645B} report the discovery of Sedna, the first of a number of objects of the inner Oort cloud to be detected in subsequent years. 
The discovery of 2012~VP$_{113}$ \citep{2014Natur.507..471T} leads the authors to conclude that a perturber at 250\,au could explain the apparent clustering in the arguments of perihelion of the distant scattered disk population around $\omega\approx0$\degr, an effect also observed in subsequently discovered objects \citep{2016AJ....152..221S,2017AJ....153..262B}. The underlying simulations were, however, limited, and the authors stress that the properties of the potential perturber are not unique.   
\cite{2016AJ....151...22B} not only show that this clustering extends to physical space (i.e. orbital planes), but that it can be explained by the existence of a ninth planet (referred to here as Planet Nine) in our solar system with a mass of $\approx10\,M_{\oplus}$ on an orbit with a $\approx 700$\, au semimajor axis.
This publication has sparked a vivid debate on the existence of Planet Nine, leading to publications which support the hypothesis \citep[e.g.][]{2016AJ....152..126B,2016AJ....152..215L,2016ApJ...833L...3B,2017AJ....154...65B,2017AJ....153...27G,2017AJ....154...61B,2017MNRAS.471L..61D,2017AJ....154..229B}, as well as those that argue against it \citep[e.g.][]{2017AJ....153...33L,2017ApJ...845...27N,2017AJ....154...50S}.
Furthermore, other authors derive low probabilities for the production of the proposed orbit \citep{2016ApJ...823L...3L}.
The fact that Planet Nine has not been detected to date must then be due to one of the following premises: {\it (i)} It does not exist; {\it (ii)} previous surveys lack the necessary sensitivity; or {\it (iii)} the object's signature is present in survey data but has hitherto eluded all applied detection algorithms.

Several authors provide restrictions on the location of Planet Nine based on {\it Cassini} data \citep{2016A&A...587L...8F,2016AJ....152...94H}, motion of comets \citep{2017AstL...43..120M}, a Monte Carlo approach \citep{2016MNRAS.459L..66D}, astrometry of Trans-Neptunian Objects (TNOs) \citep{2016AJ....152...80H}, mean motion resonance \citep{2016ApJ...824L..22M,2017AJ....153...91M} and the sky coverage of sufficiently deep surveys \citep{2016arXiv160305712B}.
Despite these spatial constraints, the search for an object as distant as the proposed Planet Nine illustrates the trade-off between sky coverage and limiting magnitude. Brightness estimates \citep{2016ApJ...824L..25F,2016A&A...592A..86T,2016arXiv160207465L} place it at the edge of the detection limit of most large surveys, while a dedicated search with large telescopes is relatively expensive in terms of observing time.

However, Planet Nine may be hidden within existing survey data. A variety of methods have been proposed to discover faint, moving objects within these vast data sets. For example, \cite{2017ApJ...841L..19K} have developed a citizen science project in which volunteers visually check images acquired by the {\it Wide-Field Infrared Survey Explorer} (WISE, \cite{2010AJ....140.1868W}), and \cite{2017arXiv171204950M} use time-resolved co-adds to rule out the existence of Planet Nine within the WISE data up to a W1 magnitude of $16.7$. 
Despite differing in the precise implementation, most other methods for the detection of distant TNOs \citep[e.g.][]{2004MNRAS.347..471P,2013PASP..125..357D,2015AJ....149...69B,2016arXiv160704895W} are based on the same principle: data acquired during a time span of several days is utilized to identify {\it tracklets}, that is, moving sources. Tracklets from different epochs are then cross-checked via an algorithm based on or similar to the work of \cite{Bernstein2000}. All of these 
algorithms seem to require multiple tracklets within a time frame of $\approx60$ days.
In a recent publication, \citep{2018ApJ...855L...6H} describe the discovery of a TNO with the Pan-STARRS-1 Outer Solar System Pipeline \citep{2011AAS...21743502H}, which transforms single detections from topo- to heliocentric coordinates and subsequently searches for linearity. However, in this approach the heliocentric distance is not a fit parameter, but rather looped through step-wise.

While having successfully detected TNOs, all previous approaches based on tracklets rely on the existence of several exposures acquired within a short span of time. In order to query data in areas with low coverage depth and/or contamination through various effects, it is possible to implement an algorithm which corrects for the objects' parallactic motion and which is then able to correlate sparsely sampled data acquired over long time intervals.

In this publication, we present our new approach to moving object detection, which we then employ to reexamine the WISE/NEOWISE single-exposure source catalogs. Throughout this publication, we use the TNO (136472) Makemake \citep{2005IAUC.8577....1B} to illustrate the challenge of moving object detection and as a proof-of-concept of our method. Our paper is structured in the following way: In section\,\ref{sec:newapproach} we describe the concept of the algorithm along with a mathematical description of the steps involved. In order to demonstrate functionality, we test the algorithm on the field containing Makemake in section\,\ref{section:wisecats}, followed by an all-sky search for Planet Nine and injected artificial test planets.
We conclude with a summary and outlook in section~\ref{sec:summary}.

\section{A new approach to moving-object detection}
\label{sec:newapproach}
Typical photometric catalogs contain millions or even billions of entries (e.g., the NEOWISE single-exposure source table alone comprises a total of $\approx6\times10^{10}$ entries). The goal of our new moving-object detection approach is to correlate those singular point-source detections belonging to one physical object, without any boundary conditions on scan frequency and temporal spacing, using a specifically optimized algorithm in order to detect distant moving objects in such large data sets.

Our algorithm combines routines that filter, correlate, and select moving-object candidates with capabilities to handle solar system object (SSO) trajectories that are contaminated by random background stars.
The approach can be divided into four steps, of which we give a summary here and which we explain in detail in sections\,\ref{sec:datasel}-\ref{sec:backtracing}.
\begin{enumerate}
 \item {\it Data selection and filtering:} Choose the field of interest and remove all data with temporal duplicity (i.e., background stars).
 \item {\it Clustering:} Form groups of single-observation point sources which could originate in real moving objects.
 \item {\it Selecting:} Compute all possible permutations of three point sources out of the clustered groups and iteratively try to linearize these subgroups via transformation to heliocentric coordinates; assume that in the time span of several years the orbit is linear in these coordinates; compile a set of subgroups with linearity in heliocentric coordinates at a corresponding heliocentric distance.
 \item {\it Backtracing and position prediction:} All resulting subgroups are re-clustered and selected, this time requiring a larger subgroup size. The linearized orbit of sufficiently large subgroups can be extrapolated to future dates so as to provide a position prediction for follow-up observations.
\end{enumerate}

\noindent In the following we give a detailed description of the different steps performed by our algorithm.

\subsection{Data selection and filtering}
\label{sec:datasel}
\bef
\centering
\resizebox{0.95\hsize}{!}{\includegraphics{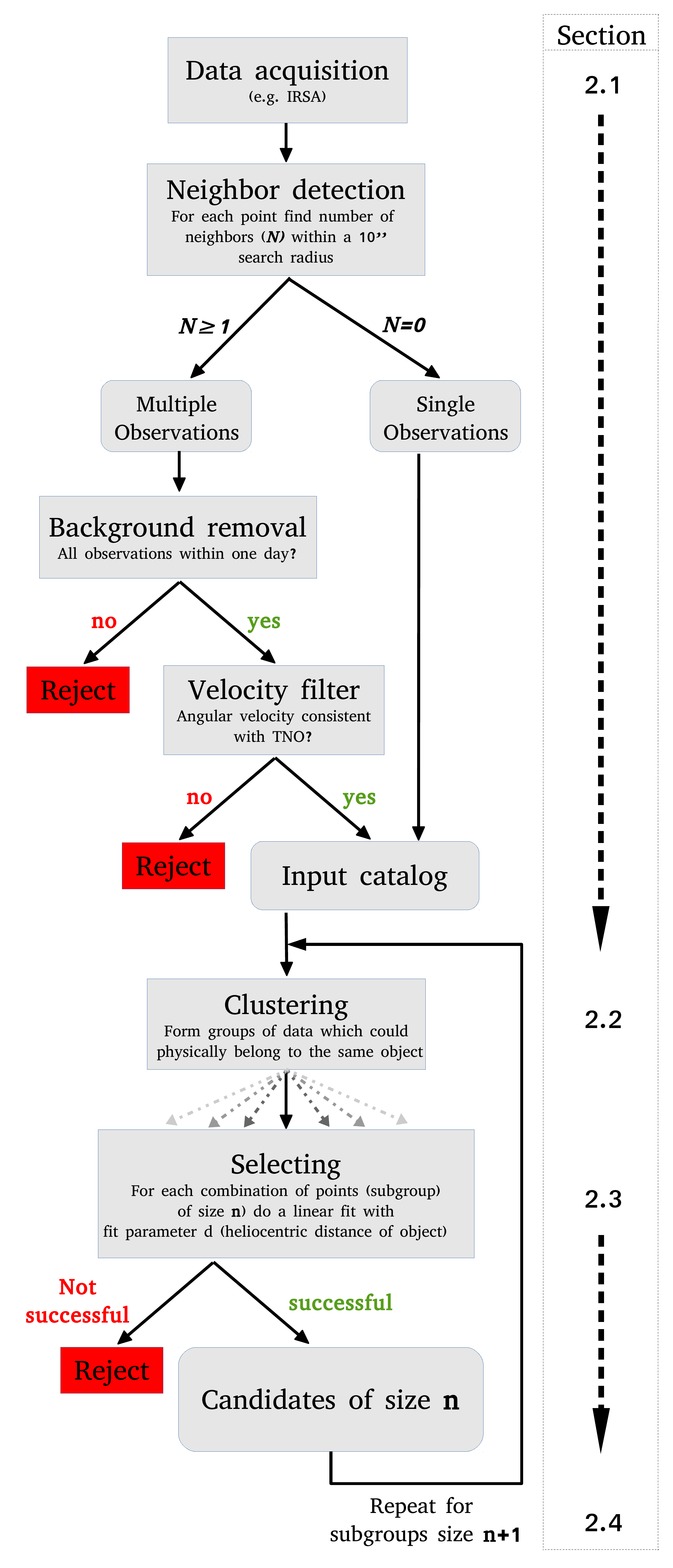}}
\caption{Flowchart of the search algorithm. All process are displayed as square boxes, while all (intermediate) catalogs are marked by rounded boxes. The corresponding section in the text is given in the right column.}
\label{fig:geo2}
\eef
Due to its spectral coverage and observing strategy, the {\it Wide-Field Infrared Survey Explorer} \citep{2010AJ....140.1868W} is well-suited for the search for distant SSOs.
Covering each area of the sky multiple times during half-year intervals,  the WISE mission provides all-sky data spanning a total of 4\,years. We therefore used WISE data for the first tests of our algorithm, but we stress that the method works with all large area/large temporal coverage survey data sets.
We downloaded single-exposure source lists from the WISE catalog via the {\it NASA/IPAC Infrared Science Archive}, requiring a separation from the {\it South Atlantic Anomaly} of more than 1 degree, no direct neighbors in the 2MASS catalog \citep{2006AJ....131.1163S, 2013yCat.2328....0C}, a detection in the W2 band ($w2mpro\neq null$), and all other quality and saturation flags with appropriate values. Furthermore, the SSO Flag ({\it sso\_flg}) in the WISE catalogs was required to be zero to avoid re-discovery of known objects.

In order to remove stationary background sources (e.g., stars and galaxies), we first checked for temporal duplicity and then filtered the data in the following manner: For each observed event we found all other data within a specific search radius. This radius is determined by two factors: {\it (i)} the maximum orbital motion of an SSO at a given distance and {\it (ii)} its maximum parallactic motion.
In order to approximate these two contributions, we followed the reasoning of \cite{2016ApJ...822L...2C} and determined the hourly orbital movement assuming a semimajor axis of 700\, au as
\begin{equation}
\mu_{orb}=10^{-2} \text{arcsec/hr} \left(\frac{d}{700 \text{au}}\right)^{-3/2}
,\end{equation}
and the hourly parallactic movement as
\begin{equation}
\mu_{par}=10^{-1} \text{arcsec/hr} \left(\frac{d}{700 \text{au}}\right)^{-1},
\end{equation}
where {\it d} denotes the distance of the object.
Throughout most of the sky, WISE covers a given region $\approx 12$ times per half year, and all visits occur within $\approx 1$ day, so the overall apparent movement becomes $\mu\approx10$\,arcsec/d, assuming a distance of $d=180$\,au. We note that, since singular points (i.e., those without neighbors within the search radius) can be carried along for further processing (as explained below), the choice of a smaller radius does not result in the loss of true events. It does, however, yield a higher number of points for further processing. The choice of a larger search radius, on the other hand, eliminates real SSO observations along with the background.\\
For those groups with two or more points, the average right ascension, declination, and Modified Julian Date (MJD) was calculated. Finally, the angular velocity of each group was computed and used as a further filtering tool.
Since the angular velocity caused by Earth's orbit is dependent on distance of the object, time, and direction, we used the {\it AstroPy} package to compute an upper and lower limit for the angular motion within one day by defining a minimum and maximum distance search window. Allowing for a margin of 1\,arcsec d$^{-1}$ to account for the orbital motion of objects, we selected only those objects exhibiting a velocity in the determined window for further processing.

\subsection{Clustering} 
\label{sec:clustering}
Using the compiled input catalog, the algorithm computes clusters of data which could physically belong to an orbit. This is done by comparing every pair of object vectors 
\begin{equation}
\vec{X_i} = \left( \alpha_i , \delta_i , t_i \right)\,,
\end{equation}
in terms of their temporal distance $\Delta t$ and angular great circle distance
\begin{equation}
\cos \Delta \rho = \sin \delta _1 \sin \delta _2 + \cos \delta _1 \cos \delta _2 \cos(|\alpha _1 - \alpha _2|)\,.
\end{equation}
A search radius is defined by assuming a proper motion and a parallax, inside of which a pair of vectors $\vec{X_i}$ is accepted to belong to a cluster.
The number of vectors $\vec{X_i}$ clustered together is then called the cluster size.\\
For a nearly geocentric observer, all TNOs show a characteristic loop whose shape is controlled by the Earth's orbit as well as the proper motion of the object and its coordinates. In order to illustrate this we used the \textit{HORIZONS Web-Interface}\footnote{\url{https://ssd.jpl.nasa.gov/horizons.cgi}} to compute an artificial sample of observations of Makemake in the period 2016 Jan 1 and 2017 Sep 1 as displayed in Fig.~\ref{fig:makemakereal1}.\\
An upper limit for the semimajor axis of the parallactic loop is given by
\begin{equation}
p _\mathrm{max} = \arcsin \left( \frac{1\,\au}{d_\mathrm{min}} \right)\,,
\end{equation} 
where $d_\mathrm{min}$ denotes the minimal geocentric distance. For a given temporal separation $\Delta t$, we can estimate an upper limit for the parallactic component of the search radius via 
\begin{equation}
p = 2 \, p _\mathrm{max} \, \sin \left( \omega _\oplus \, \Delta t \right)\,,
\end{equation}
where $\omega _\oplus = 2\piup / yr$ is the Earth's mean orbital angular velocity.\\
For distant SSOs, the parallactic loop is mildly distorted by a proper motion component. Imposing a \textit{minimum} semimajor axis and a \textit{minimum} distance $d_\mathrm{min}$, we can calculate a \textit{maximum} perihelion orbital velocity
\begin{equation}
v_\mathrm{max} = \sqrt{ G \, \Msol \, \left( \frac{2}{d_\mathrm{min}}-\frac{1}{a_\mathrm{min}} \right)}\,,
\end{equation} 
where we approximate the object's mass to be zero. With this peak velocity, we can calculate a maximum daily proper motion component via
\begin{equation}
m_\mathrm{max} = \arctan \left( \frac{v_\mathrm{max}\, 1\,\mathrm{d}}{d_\mathrm{min}} \right)\,.
\end{equation}
Finally, the total search radius is now determined via
\begin{equation}
\label{eq:max_distance}
r_\mathrm{max} = m_\mathrm{max} \frac{\Delta t}{1\,\rm{d}} + \, 2 \, p _\mathrm{max} \, \sin \left( \omega _\oplus \, \Delta t \right)\,.
\end{equation} 
The clustering algorithm was implemented in the \texttt{FORTRAN} programming language and fully parallelized for shared-memory machines using \texttt{OpenMP}. Even large sets of data containing millions of objects can be clustered within hours or a few days. An example of a typical cluster size distribution is given in Sect.\,\ref{sec:40-15}. For this example, the required computing time on a single {\it Intel\textsuperscript{\textregistered} Xeon\textsuperscript{\textregistered} CPU E5-2680 v3 @ 2.50GHz} was $\approx20$\,h.

\subsection{Selecting}
\label{sec:select}
Even for non-SSOs, we have a non-zero probability of two objects having just the right temporal and angular separation to be regarded as part of the path of one virtual new SSO.
Thus, we established a filtering pipeline that selects the physically plausible clusters.
To do so, all clusters are transformed from their initially nearly geocentric reference system into a heliocentric system assuming a distance $d$.
As a result, paths of actual distant SSOs are to a good approximation linear in the $(\alpha_{\odot},\delta_{\odot})$ plane over long timescales if $d$ is chosen correctly.
Figure~\ref{fig:sso1} shows the same observations as Fig.~\ref{fig:makemakereal1}, but transformed into heliocentric coordinates.
Therefore, the core idea of our algorithm is to assume that a given cluster represents the observed path of an SSO within some search distance range $[d_\mathrm{min},d_\mathrm{max}]$.
Subsequently, the algorithm tries to linearize the cluster by applying subsequent transformations to the heliocentric system and finding the best-fit distance.

\subsubsection{Linearization}
\label{sec:linearity}
For the sake of simplicity and to reduce the computational effort, we apply two simplifications. First, we perform no barycentric corrections leading to peak errors on the order of $\leq0.01$\,au. Furthermore, we assume that the SSO's heliocentric distance does not change significantly throughout the measurements. Depending on the heliocentric distance and the time range of the cluster, additional uncertainties up to $\approx0.2$\,au can occur. Both simplifications are easily met by distant Kuiper belt objects and lead to total positional errors of just a few percent (assuming heliocentric ranges greater than $30$~au).\\
We consider a heliocentric coordinate system and the x-axis pointing towards the vernal equinox.
\newcommand\Bigcircle{\raisebox{-0.5mm}{\scalebox{1.7}{$\bigcircle$}}}

\bef
\centering
\resizebox{0.6\hsize}{!}{\includegraphics{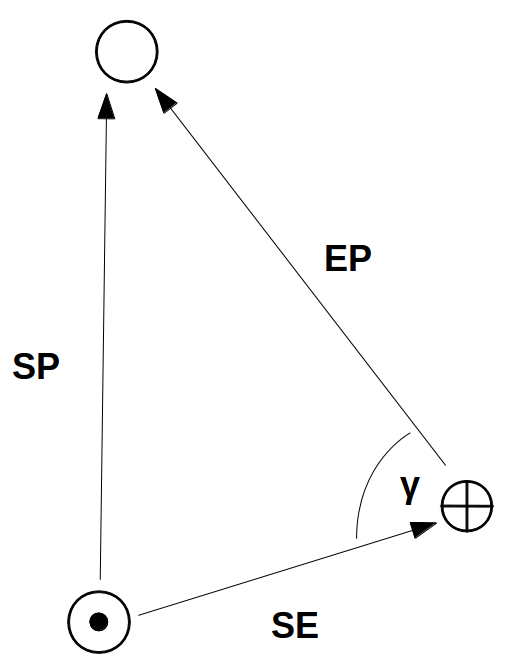}}
\caption{Schematic of the vectors introduced in \ref{sec:linearity}. The Sun is denoted by $\odot$, Earth by $\oplus$ and the TNO by $\bigcirc$.}
\label{fig:geo}
\eef

\bef
\resizebox{\hsize}{!}{\includegraphics{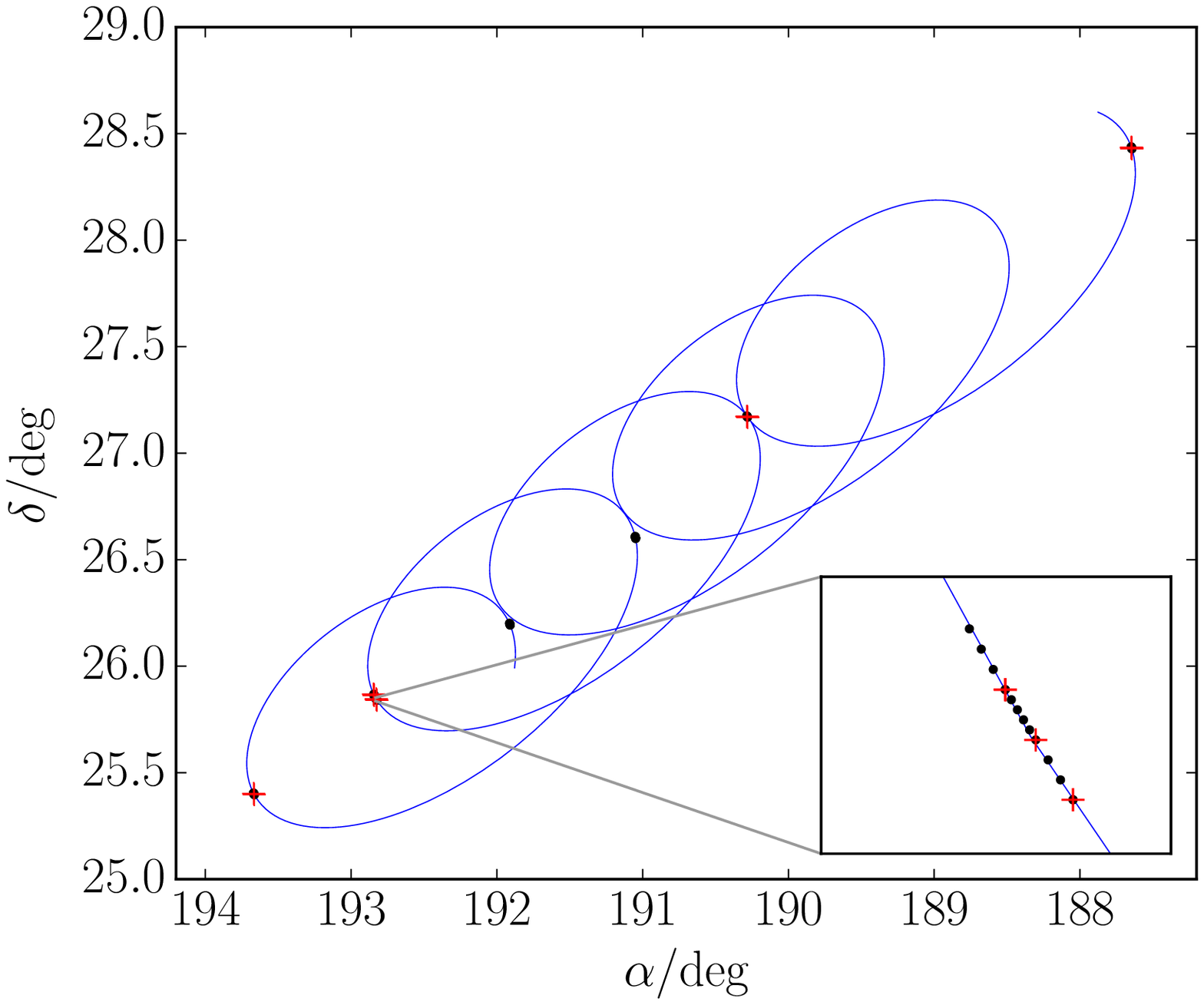}}
\caption{Makemake geocentric path between 2010 May 13 and 2015 April 7 generated using the \textit{HORIZONS Web-Interface} (blue line). All WISE observations during which Makemake was in the field of view are marked with a dot. The WISE detections of Makemake are marked with a red cross.}
\label{fig:makemakereal1}
\eef

\noindent We take $\vec{SE}$ to be the vector pointing from the Sun towards Earth, $\vec{EP}$ the vector pointing from Earth towards the SSO candidate and $\vec{SP} = \left( x_\text{p}, y_\text{p}, z_\text{p} \right)^T$ the vector pointing from the Sun towards the SSO candidate (see Fig.~\ref{fig:geo} for a visualization). We make use of the IAU SOFA library\footnote{\url{http://www.iausofa.org}} for the calculation of the Sun-Earth position vector. 
Further,  we take $\alpha$ and $\delta$ to be the geocentric equatorial coordinates of the SSO candidate and $\gamma$ to be the angle between $\vec{SE}$ and $\vec{EP}$. The Earth-SSO vector can be written as $\vec{EP} = \Delta \cdot \vec{e}_\mathrm{EP}$, denoting the Earth-SSO distance as $\Delta$ and $\vec{e}_\mathrm{EP} = \left( \cos \delta \cos \alpha, \cos \delta \cos \alpha, \sin \alpha \right)^{T}$ being the unit vector of $\vec{EP}$. Finally, we take $d$ to be the distance between the Sun and the SSO. From this, we calculate $\gamma$ as

\bef
\resizebox{\hsize}{!}{\includegraphics{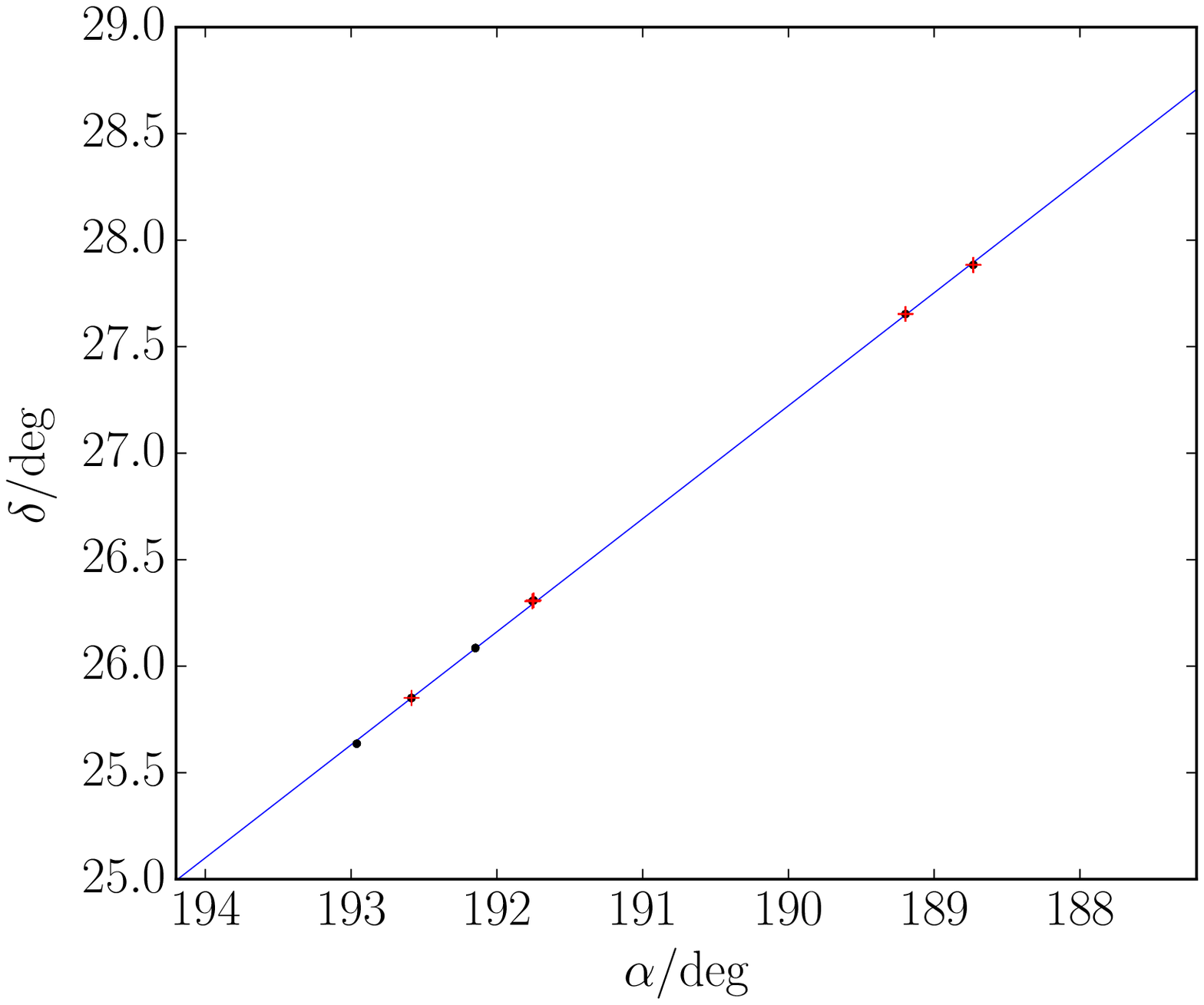}}
\caption{Illustration of the Makemake orbit in heliocentric coordinates. Transformation to heliocentric coordinates has removed the ellipsoidal component of the orbit and the epochal sets of observations (chronologically from upper right to lower left) are linearized to good approximation. All WISE observations during which Makemake was in the field of view are marked with a dot. The WISE detections of Makemake are marked with a red cross.}
\label{fig:sso1}
\eef

\begin{equation}
\cos \gamma = -\frac{\vec{SE} \cdot \vec{EP}}{|\vec{SE}| \cdot |\vec{EP}|} = -\vec{e}_\mathrm{SE} \cdot \vec{e}_\mathrm{EP}
,\end{equation}
with Sun-Earth unit vector $\vec{e}_\mathrm{SE}$ and the negative sign to make sure that we calculate the inner angle between the two vectors.\\

Now, according to the law of cosines, the identity
\begin{equation}
|\vec{SP}|^2 = |\vec{SE}|^2 + |\vec{EP}|^2 - 2 \, |\vec{SE}| \, |\vec{EP}| \, \cos \gamma 
\end{equation}
holds for the triangle spanned by the three vectors, which is equivalent to
\begin{equation}
d^2 = r^2_\oplus + \Delta ^2 - 2 \, r_\oplus \, \Delta \, \cos \gamma
,\end{equation}
with the Sun-Earth distance $r_\oplus$. Solutions for $\Delta$ are given by
\begin{equation}
\Delta = r_\oplus \left( \cos \gamma \pm \sqrt{\cos^2 \gamma + (d/r_\oplus)^2-1} \right),
\end{equation}
but as we are interested in objects beyond Earth's orbit, only the positive solution is relevant. This equation implies that for a given guess of the heliocentric distance $d$, the geocentric distance can be calculated, and thus the full Sun-SSO vector via $\vec{SP} = \vec{SE} + \Delta \, \vec{e}_\mathrm{EP}$. Given that vector, we can calculate heliocentric equatorial coordinates $\alpha _\sun$ and $\delta _\sun$ using
\begin{equation}
\alpha _\sun = \arctan \left( \frac{y_\mathrm{p}}{x_\mathrm{p}} \right) \comequa \, \, \, \delta _\sun = -\arctan \left( \frac{z_\mathrm{p}}{x_\mathrm{p}^2+y_\mathrm{p}^2} \right) \fsequa
\end{equation}

In these coordinates, any set of vectors of equatorial positions for some given time $t_i$
\begin{equation}
\vec{X}_i = \left( \alpha_i , \delta_i , t_i \right) 
,\end{equation}
can be transformed into a set of vectors of heliocentric positions 
\begin{equation}
\vec{X}_{\sun,i} = \left( \alpha_{\sun,i} , \delta_{\sun,i} , t_i \right) 
,\end{equation}
for a given heliocentric distance guess $d$. Random sets of vectors $\vec{X}_i$ are not expected to show any special behavior in the $(\alpha _\sun,\delta _\sun)$ plane, while those sets of vectors $\vec{X}_i$ belonging to an SSO are transformed into a line in the $(\alpha _\sun,\delta _\sun)$ plane if the provided distance guess $d$ proves to be consistent with the object's true heliocentric distance. The quality of the best-match heliocentric range is estimated by calculating the linear Pearson correlation coefficient for the set of $n$ vectors $\vec{X}_{\sun,i}$ through
\begin{equation}
\rho_{\alpha\delta} = \frac{\sum \alpha_{\sun,i} \delta_{\sun,i} - n \bar{\alpha}_\sun \bar{\delta}_\sun}{\sqrt{\sum \alpha_{\sun,i} ^{2} - n \bar{\alpha}_\sun^{2}} \sqrt{\sum \delta_{\sun,i} ^{2} - n \bar{\delta}_\sun^{2}}} \fsequa
\end{equation}
Actual SSOs with a large semimajor axis are transformed into lines in the $(\alpha _\sun,\delta _\sun)$ plane with correlation coefficients close to unity. Large samples of data covering sufficiently large temporal intervals lead to well-defined maxima of the squared Pearson correlation coefficient $\rho_{\alpha\delta}^2$ which can be calculated via numerical maximization algorithms. Figure \ref{fig:makemake_correlation_map} shows $\rho_{\alpha\delta}^2$ as a function of heliocentric distance for a sample of Makemake data sets. Notably, the best-fit heliocentric distance coincides with the mean heliocentric distance of the object for the given time span.

\bef
\resizebox{\hsize}{!}{\includegraphics{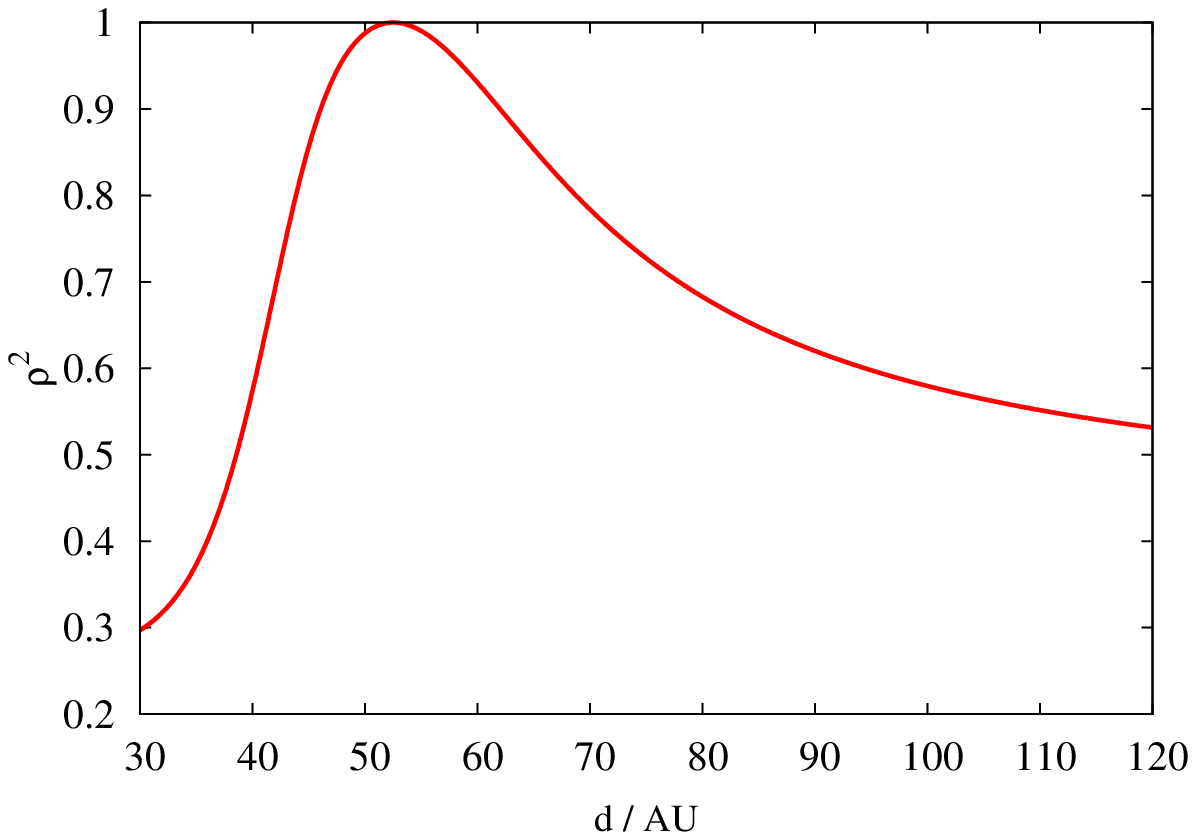}}
\caption{Squared Pearson correlation coefficient $\rho^2$ as a function of the heliocentric distance guess $d$ for a set of 37 calculated geocentric positions of Makemake between 01-01-2017 and 01-01-2018. The function peaks prominently at 52.5\, au which is the average heliocentric distance of Makemake for the given time interval.}
\label{fig:makemake_correlation_map}
\eef

\noindent Very small samples ($\approx$3) covering small temporal intervals (less than a month) feature less prominent $\rho ^2$ peaks and show a strong dependence on the observation time. For instance, during quadrature the apparent motion takes a minimum. As a result, we obtain smaller geocentric arcs, less well defined $\rho ^2$ peaks and more complicated $\rho ^2$ functions in general, which may result in incorrect best-fit distances if the provided fitting range is too large (see Fig.~\ref{fig:makemake_correlation_difficult}). Therefore, we apply a number of post-fit sanity checks to detect failed linearizations (see Sect.\,\ref{sec:subcluster_analysis}).

\bef
\resizebox{\hsize}{!}{\includegraphics{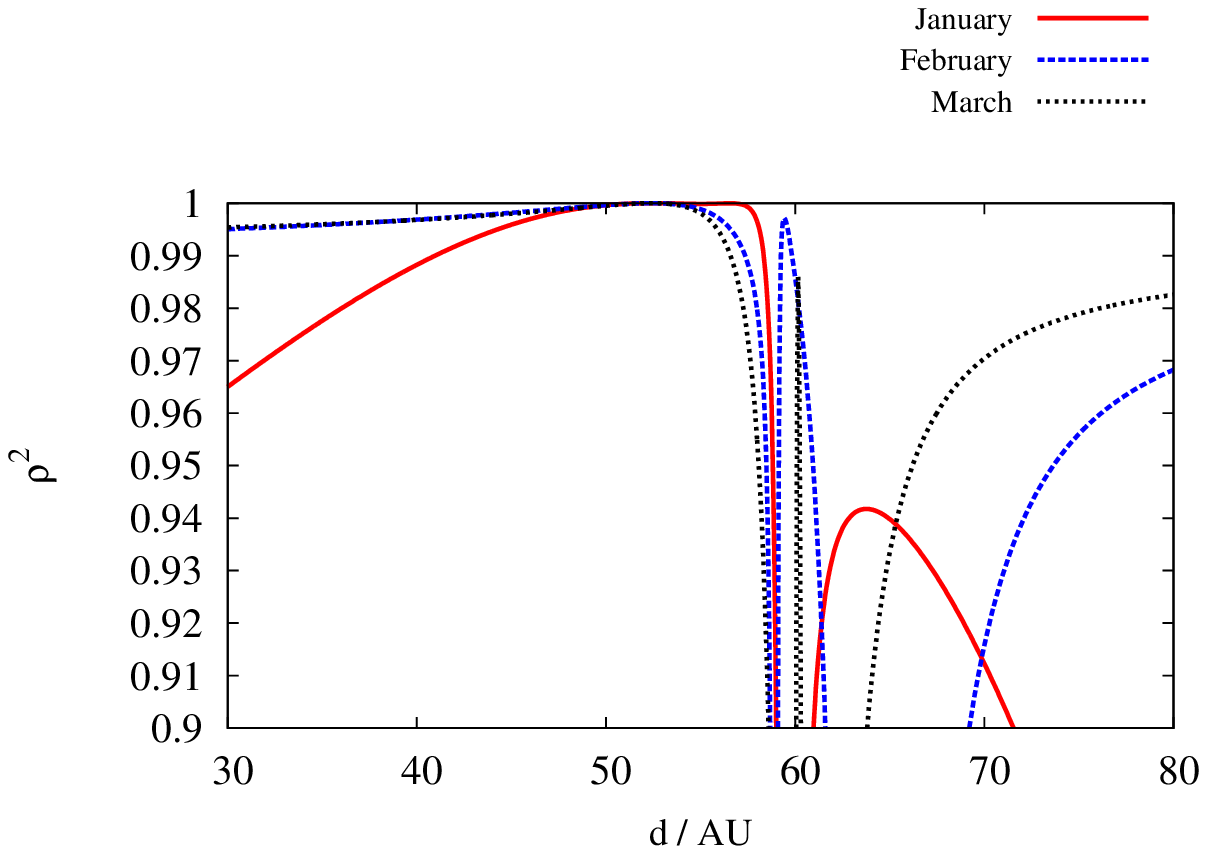}}
\caption{Squared Pearson correlation coefficients as functions of the heliocentric distance guess for different samples (1 day cadence) of Makemake ephemeris, all covering 1 month in 2017. Despite almost identical time spans, numbers of observations and observation frequency, the $\rho ^2$ functions differ greatly. For such data, the fitting range $[d_\mathrm{min}, d_\mathrm{max}]$ must closely bracket the actual physical peak. Otherwise, the maximization routine may end up stuck at the borders of the fitting range or at one of the secondary peaks. While this could be avoided by checking linearity for small steps in heliocentric distance, this would void the speed of the fitting routine; so instead, the analysis is carried out with steps of $d_{max}-d_{min}=10$\,au.}
\label{fig:makemake_correlation_difficult}
\eef

\noindent We conclude that our linearization algorithm works best for larger group sizes with a minimum of three data points covering a sufficiently long temporal range of ten days or more. Larger time spans and subgroup sizes increase the quality of the fit. For very small time spans, higher sampling rates are necessary in order to reproduce at least a small fraction of the parallactic loop and obtain the correct distance. However, tools such as MOPS cover that parameter range and are a natural complement to our approach. Compared to full orbital fitting routines which typically minimize parameter vectors with six fit parameters \citep[see, e.g.,][]{Bernstein2000}, our simplified approach deals with only {one} single fit parameter (i.e., the heliocentric distance $d$) to identify SSO candidates, and allows us to process much larger sets of data.
\subsubsection{Subcluster analysis}
\label{sec:subcluster_analysis}
Given the non-zero probability of cluster contaminations through nearby stars, we perform a subcluster decomposition for every SSO candidate cluster identified by our pipeline. Explicitly, for every cluster of size $n$ we consider every possible permutation of subsets with subgroup size $k$ leading to a total number of sub-clusters per cluster of
\begin{equation}
N_\mathrm{sub} = \binom{n}{k}
,\end{equation} 
ensuring a piecewise direct selection of uncontaminated planetary trails. For all of our first runs, we choose a subgroup size of $k=3$ or $k=4$.
All subclusters are transformed into the best-fit heliocentric distance and tested for linearity as described in Sect.\,\ref{sec:linearity}. Because of the large number of subclusters per cluster (e.g., 1140 for $n=20$ and $k=3$), we apply additional constraints to reduce the number of false positives. First, we require a proper time sequence of the heliocentric coordinates. Secondly, we check if the subgroup path length is consistent with the proper motion of an SSO. Finally, recalling that actual planetary paths show very high linearity in heliocentric coordinates, we require a squared Pearson correlation coefficient of $\rho_{\alpha\delta}^{2} \geq 0.9999$.
\bef
\resizebox{\hsize}{!}{\includegraphics{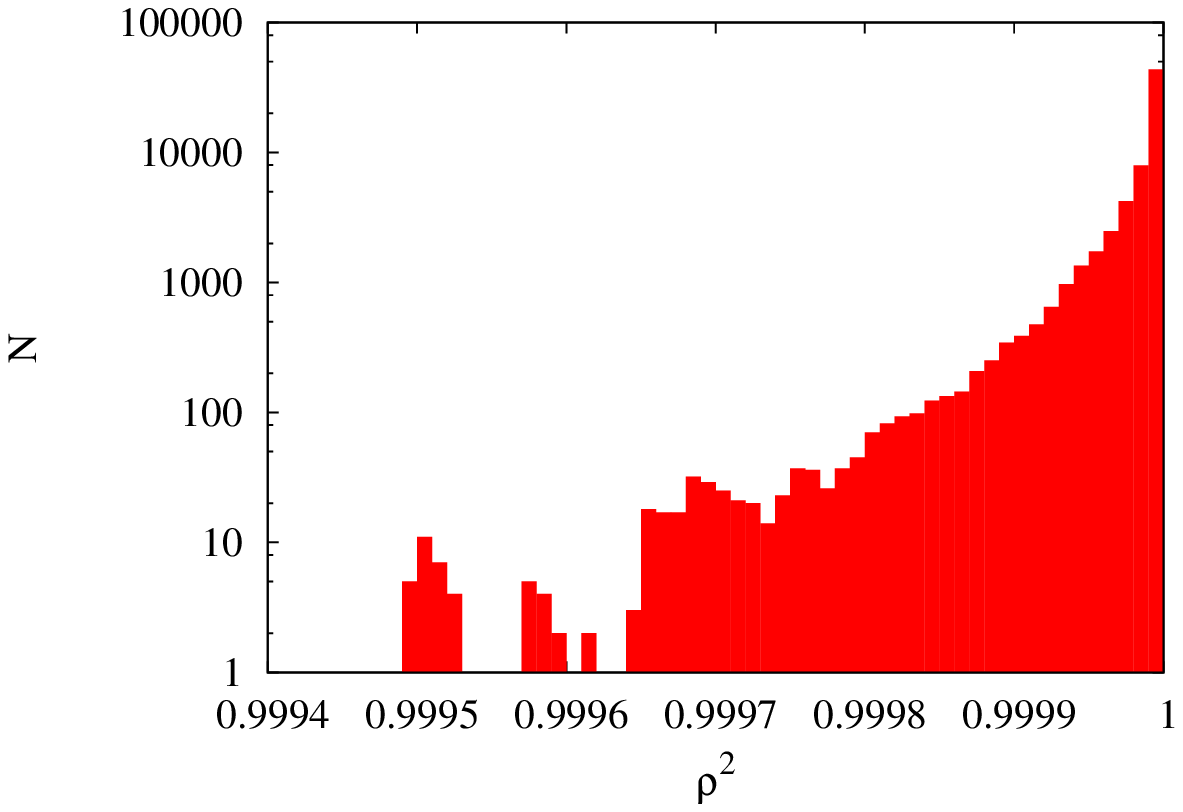}}
\caption{Distribution of the squared Pearson correlation coefficients of all subgroups of size $k=4$ for the Makemake data cluster presented in Fig.~\ref{fig:makemake_correlation_map}. The vast majority of subgroups are well-fitted and yield good linearity values larger than $0.9999$.}
\label{fig:makemake_rho_dist}
\eef
In Fig.~\ref{fig:makemake_rho_dist} we applied a subcluster decomposition to the Makemake data cluster shown in Fig.~\ref{fig:makemake_correlation_map} with a subgroup size of $k=4$, resulting in subclusters spanning between $40$ and $360$ days, and calculated the distribution of the squared Pearson correlation coefficients. The distribution peaks at values larger than $0.9999$.
\subsection{Backtracing and position prediction}
\label{sec:backtracing}
After the subcluster analysis, we are typically (for the whole WISE catalog) left with $\approx 100$ subgroups/deg$^2$, every one of which consists of the coordinates and time of observation as well as the best-fit distance. 
As the probability of producing subgroups of size three with a high degree of linearity from random data (noise) is non-zero, all data from the previous step (Sect.\,\ref{sec:subcluster_analysis}) are then re-clustered and selected, this time requiring the same squared Pearson correlation, but with a larger subgroup size. Since every group of four must consist of groups of three, all valid candidates are preserved while eliminating random matches. This process can be repeated until the required subgroup size deemed a ``real'' candidate is reached.
In order to be able to carry out follow-up observations, the linearized (i.e., heliocentric) orbit is then extrapolated to the observation date and transformed to geocentric coordinates.

\section{Results}
\label{section:wisecats}
\begin{figure*}[ht!]
\centering
   \includegraphics[width=\textwidth]{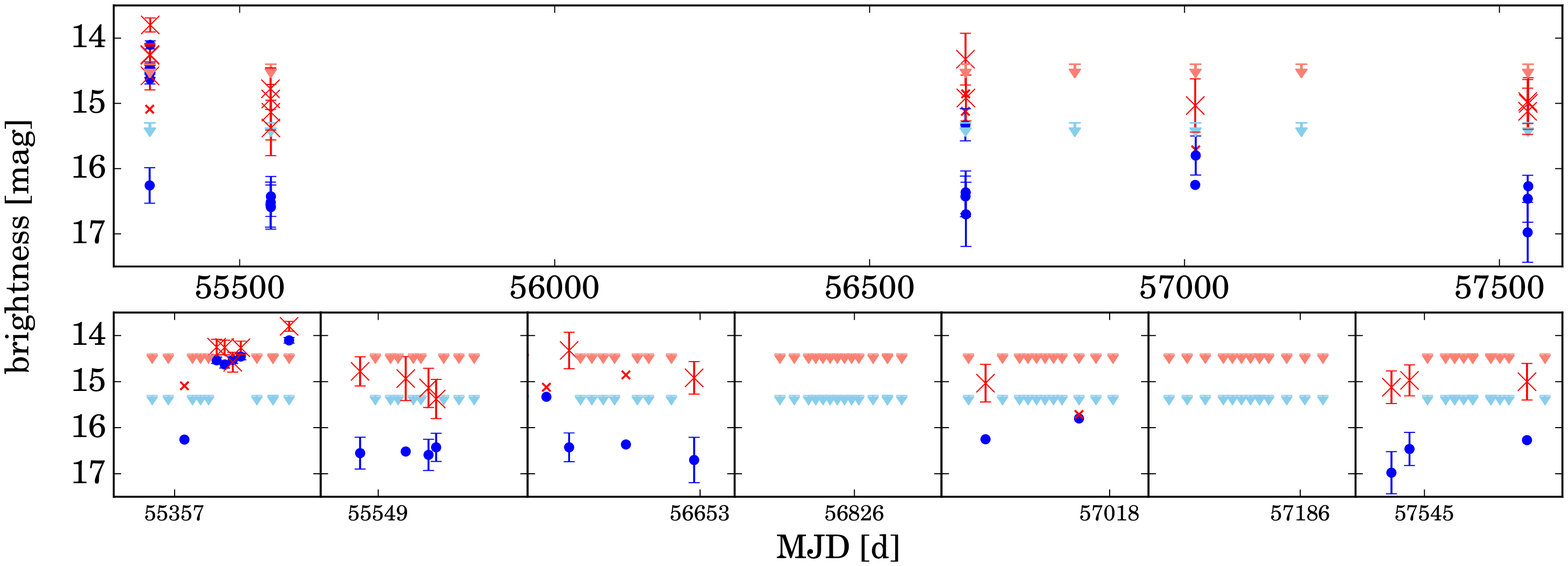}
     \caption{Light curve of Makemake in the WISE bands W1 (red crosses) and W2 (blue dots). The upper row shows the entire light curve, while the second row shows a zoom of the observation epochs in 1.5\,d windows. Non-detections are marked with an arrow at the S/N=5 level. We note that most detections have a S/N below 5.}
     \label{mm_lc}
\end{figure*}

\subsection{Makemake}
\label{sec:makemake}
As a proof-of-concept we tested our algorithm by performing a blind detection of Makemake \citep{2005IAUC.8577....1B}, that is, with no prior assumptions. 
The {\it Known Solar System Association Table} of the NEOWISE mission denotes a total of 32 detections of Makemake, which were attributed to it {\it a posteriori}. However, the SSO was in the field-of-view a total of 124 times, meaning that it was not detected in the majority of exposures. This can be explained by the  fact that, with a mean brightness in band W1 and W2  of $\overline{w1}=16$\,mag and $\overline{w2}=14.9$\,mag, Makemake is at the detection limit of WISE single exposures. Furthermore, it exhibits a large variability in brightness in both bands, with a range of 3\,mag in W1 and 2\,mag in W2, caused by effects such as rotational modulation, changes in background and noise. Figure~\ref{mm_lc} shows the light curve of Makemake in the W1 and W2 passbands for the entire WISE data set released so far, along with non-detections.
It therefore presents a good example to illustrate the difficulty of discovering such objects within catalog data via standard methods, such as the search for ``tracklets'', as well as a test for the method described in this publication.

We downloaded WISE data covering the Makemake orbit during the period 2010-2016 (i.e., $\alpha=187.5$ to $194$, $\delta=+25$ to $+29$) with the settings described in section\,\ref{sec:datasel}, producing a total of 280\,515 point sources\footnote{We note that this section of the sky was not covered during the WISE 3-band mission phase.}. Since the {\it sso\_flg} was set to zero in the standard data acquisition, all WISE detections of Makemake with a detection and error estimate in {\it both} bands (a total of 16 individual detections) were added manually at this stage, that is, before any data reduction, thereby ensuring a representative trial. 
The selection and velocity filtering yielded a total of 4472 groups consisting of two single detections, and 918 groups consisting of more than two singular detections. Therefore, after computing the mean coordinates and date, we were left with 5390 points in the input catalog. After applying the velocity filter described in Sect.\,\ref{sec:datasel}, this number was reduced to 4836, which constituted our input catalog for the the clustering and selection processes.

Running the processes described in Sects.\,\ref{sec:clustering} and \ref{sec:select} with a subgroup size of three, we used the selected subgroups as input for a second run with subgroup size 4, which left a singular candidate with a Pearson coefficient of $\rho^2=0.999999999$ at a best-fit heliocentric distance of 52.6\,au. A lookup in the original data and comparison to the {\it Known Solar System Object Possible Association List} shows that the subgroup constitutes seven singular detections of Makemake. As a trace-back reveals, the remaining detections were discarded during the data selection process due to proximity to background sources. 
However, transforming the entire input catalog to heliocentric coordinates assuming the derived best-fit distance and identifying all sources which fit the linearity in $\alpha_{\odot}$ and $\delta_{\odot}$ as a function of time to within $2 ''$, we retrieved all detections of Makemake from the data set. We therefore conclude that our approach allows for the discovery of Makemake based on WISE data alone.

\subsection{All-sky search for Planet Nine and blind test}
\label{sec:40-15}
As described in section~\ref{sec:intro}, there are several, sometimes conflicting, position predictions for Planet Nine. For this reason, we did not limit our search to a specific region of the sky, but rather perform an all-sky analysis. We obtained WISE data of this region with the flag settings described in Sect.~\ref{sec:datasel} and without any {\it a priori} assumptions regarding colors or brightness. In this case, however, we limited our input data to the NEOWISE mission phase, since the presence of the {\it AllWISE flag} within this catalog allowed us to discard a large fraction of the data from the beginning. The query yielded a total of $\approx 4.2\times10^7$ individual point sources.
In Fig.~\ref{fig:density_all} we illustrate the given source density per square degree.
Using the same settings as described in Sect.~\ref{sec:datasel}, the removal of stationary sources resulted in $\approx 25\times10^6$ data points.

To ensure that our algorithm is capable of finding SSOs at distances proposed for Planet Nine, we carried out blind tests very similar to that described in \ref{sec:makemake}.
The basic idea was to generate artificial Planet Nine data points which in principle could have been observed by WISE.
To this end, we randomly generated $10^3$ sets of Keplerian orbital elements which included values for the semimajor axis $a$, the numerical eccentricity $e$, the orbital inclination $i$, the longitude of the ascending node $\Omega$, the argument of the perihelion $\omega$, and the mean anomaly $M$.
The limits for these values were chosen according to the parameters predicted by \cite{2016arXiv160305712B}, and the random values of the mentioned parameters are uniformly distributed within these limits.
To determine the ephemerides of these artificial Planet Nine-like objects we used the publically available software tool \texttt{PyEphem}\footnote{\url{http://rhodesmill.org/pyephem/}}.
We evaluated the ephemerides for all time stamps available in the NEOWISE survey frame meta data table, resulting in $7.7 \cdot 10^6$ data points for each simulated object, and cross-matched them using the frame-corner coordinates, thus excluding all data not covered by simultaneous WISE pointings, yielding approximately 290 data points per object as observed by WISE.
From these, we decided to randomly remove further data points to account for observations at the detection limit, blends, or spikes caused by bright background stars, as well as cosmic ray contamination.
The number of deleted data points was randomly determined with the goal to achieve artificial object data sets ranging almost uniformly from 3 to 20 data points.
Our artificial data set now constitutes $11\,315$ data points of 1000 Planet Nine-like objects in total.
In Fig.~\ref{fig:density_all} we show these data points in comparison to the density of data from the WISE catalog.

Finally, we injected the candidates into the real NEOWISE data. 
Based on the experience with the recovery of Makemake described in section~\ref{sec:makemake}, we used a minimum required Pearson coefficient of $\rho^2>0.999999$ (the minimum linearity of all subgroups of Makemake) and a minimum heliocentric distance of 180~au, resulting in a total of $4.7\times10^5$ subgroups. 
\bef
\resizebox{\hsize}{!}{\includegraphics{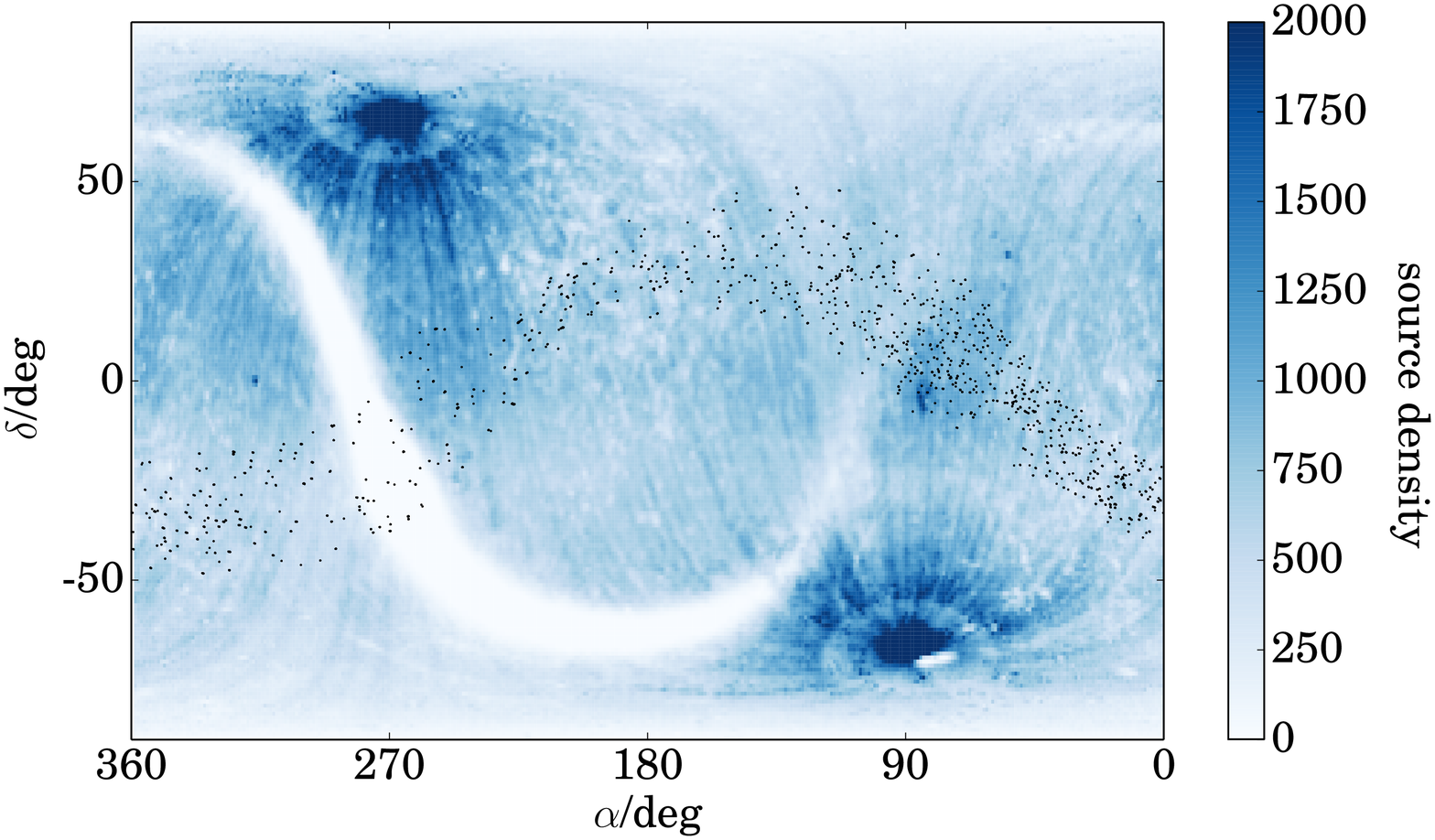}}
\caption{Density of detected sources (sources/deg$^2$) of the input catalog. For clarity, the color code has been truncated at densities above 2000, since the regions at the ecliptic poles exhibit source densities almost two orders of magnitude above the average. The injected simulated Planet Nine observations are overplotted as black dots.}
\label{fig:density_all}
\eef
In order to demonstrate the influence of point source density on the total computing time, we show the number of clusters as a function of cluster size as a solid line in Fig.~\ref{fig:density_before}. The total number of linearizations, and therefore computing time, scales with the product of the binomial coefficient and the number of clusters of a given size (dotted line). We note that the number of iterations converges toward a constant value, indicating that the processing time is manageable.

Tests have shown that the number of subgroups of size 3 and 4 formed from random data (i.e., noise) is too large to regard each of them as candidates, so a successive analysis with subgroup sizes of 4 and 5 was carried out. We stress that it is also possible to start the analysis with a higher subgroup size, although this does require larger computing times. The resulting subgroups of size 5 were merged based on the occurrence of data within multiple groups, and our algorithm retrieved a total of 836 candidates of size 5 or more from our artificial Planet Nine candidates. Considering that 872 out of the 1000 artificial planets consisted of five or more points, our algorithm successfully retrieved 95.8\,\% of the candidates, including many with only one observation per epoch. We could not find any systematics within the orbital parameters of the injected planets not recovered by the code.

Finally, since no candidates of size 5 or more were detected within the real NEOWISE data, and we were able to recover most of the injected test planets, we exclude the presence of five or more data points of Planet Nine within the NEOWISE catalog with a $2\sigma$ confidence level.

\bef
\resizebox{\hsize}{!}{\includegraphics{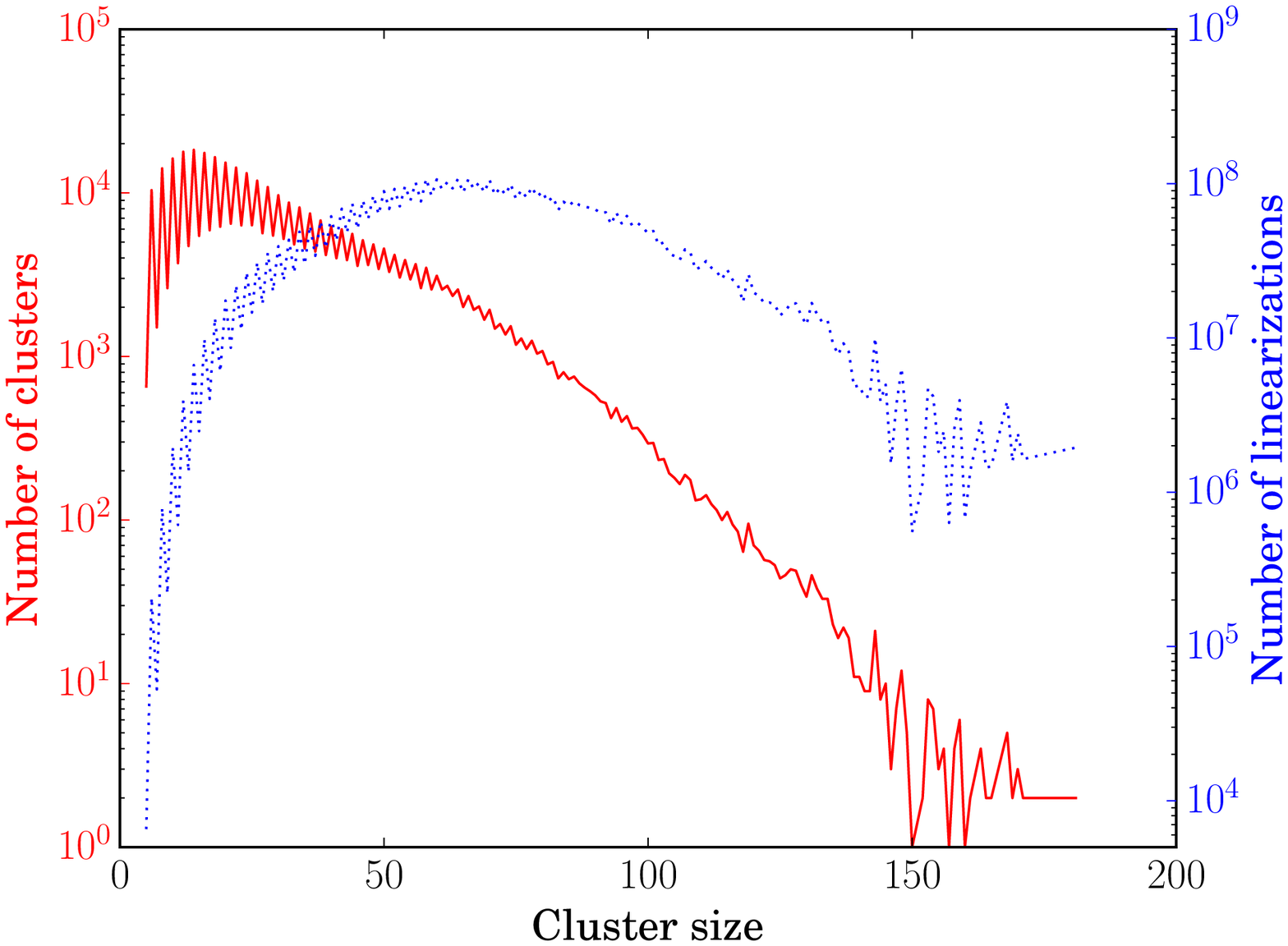}}
\caption{Number of clusters (solid line) and number of linearizations necessary to process the data (dotted line). The number of iterations converges toward $\sim 10^6$ per cluster, an achievable value with reasonable computing power.}
\label{fig:density_before}
\eef

\section{Summary and Outlook}
\label{sec:summary}
The detection of TNOs is a notoriously challenging task and requires great observational efforts.
The visual discovery of such objects via systematic observations of well defined parts of the sky has already been performed by many other groups.
We present the development of a new, highly efficient algorithm, which can deal with the unprecedented amounts of archival data that have become available within the last years.
With the additional capability of handling data covering large time spans, our code complements other existing moving-object detection algorithms.

We used WISE/NEOWISE data to test our algorithm by serendipitously detecting the known TNO Makemake.
Our code proved to be fully capable of locating objects such as Makemake in crowded fields and is accurate enough to provide position predictions for follow-up observations.

Furthermore, we demonstrated that our algorithm would have been able to detect Planet Nine in WISE data, by injecting test planets into our data, of which we were able to detect 95.8\%.

Our search for Planet Nine within WISE data was unsuccessful. 
It is, however, more likely that the object, should it exist, is not bright enough to be detected with the WISE satellite, which is in agreement with the non-detection of Planet Nine by \cite{2017arXiv171204950M}.

Having demonstrated the capabilities of the approach, we continue the search for Planet Nine by analyzing a combined data set spanning decades and comprising all large public catalogs covering the visual and infrared range. Furthermore, the fact that the code is not reliant on a tight temporal spacing of data may allow for the search for Planet Nine via stellar occultations (however rare) by compiling candidate occultation catalogs from existing photometric surveys or dedicated projects \citep{2012SPIE.8444E..0DL,2018PASP..130a4502P}.

\begin{acknowledgements}
This publication makes use of data products from the Two Micron All Sky Survey, which is a joint project of the University of Massachusetts and the Infrared Processing and Analysis Center/California Institute of Technology, funded by the National Aeronautics and Space Administration and the National Science Foundation.
This publication makes use of data products from the Wide-field Infrared Survey Explorer, which is a joint project of the University of California, Los Angeles, and the Jet Propulsion Laboratory/California Institute of Technology, funded by the National Aeronautics and Space Administration. This research has made use of the NASA/ IPAC Infrared Science Archive, which is operated by the Jet Propulsion Laboratory, California Institute of Technology, under contract with the National Aeronautics and Space Administration. The authors thank J.H.M.M. Schmitt for suggestions and proof-reading.
\end{acknowledgements}

\bibliography{astro.bib}
\bibliographystyle{aa}

\end{document}